\documentclass{article}

\usepackage{PRIMEarxiv}
\DeclareUnicodeCharacter{2212}{-}
\DeclareUnicodeCharacter{03B2}{\ensuremath{\beta}}
\usepackage[utf8]{inputenc} 
\usepackage[T1]{fontenc}
\usepackage{hyperref}       
\usepackage{url}            
\usepackage{booktabs}       
\usepackage{amsfonts}       
\usepackage{nicefrac}       
\usepackage{microtype}      
\usepackage{lipsum}
\usepackage{fancyhdr}       
\usepackage{graphicx}       
\graphicspath{{media/}}     
\usepackage{cite}
\usepackage{amsmath,amssymb,amsfonts}
\usepackage{balance}
\usepackage{algorithmic}
\usepackage{tabularx}
\usepackage{caption}
\usepackage{textcomp}
\usepackage{xcolor}
\usepackage[numbers]{natbib}
\def\BibTeX{{\rm B\kern-.05em{\sc i\kern-.025em b}\kern-.08em
   T\kern-.1667em\lower.7ex\hbox{E}\kern-.125emX}}

\title{LEXI: Large Language Models Experimentation Interface

}
\author{
Guy Laban$^{*}$ \\
  Department of Computer Science and Technology \\
  University of Cambridge \\
  Cambridge, UK\\
  \texttt{guy.laban@cl.cam.ac.uk} \\
   \And
  Tomer Laban \\
  \texttt{tomerlavan1@gmail.com} \\
     \And
Hatice Gunes \\
  Department of Computer Science and Technology \\
  University of Cambridge \\
  Cambridge, UK\\
  \texttt{hg410@cam.ac.uk} \\
}

\begin{document}
\maketitle

\begin{abstract}
\small{The recent developments in Large Language Models (LLM), mark a significant moment in the research and development of social interactions with artificial agents. These agents are widely deployed in a variety of settings, with potential impact on users. However, the study of social interactions with agents powered by LLM is still emerging, limited by access to the technology and to data, the absence of standardised interfaces, and challenges to establishing controlled experimental setups using the currently available business-oriented platforms. To answer these gaps, we developed LEXI, LLMs Experimentation Interface, an open-source tool enabling the deployment of artificial agents powered by LLM in social interaction behavioural experiments. Using a graphical interface, LEXI allows researchers to build agents, and deploy them in experimental setups along with forms and questionnaires while collecting interaction logs and self-reported data. 
The outcomes of usability testing indicate LEXI's broad utility, high usability and minimum mental workload requirement, with distinctive benefits observed across disciplines. A proof-of-concept study exploring the tool's efficacy in evaluating social HAIs was conducted, resulting in high-quality data.  A comparison of empathetic versus neutral agents indicated that people perceive empathetic agents as more social, and write longer and more positive messages towards them.}
\end{abstract}


\section{Introduction}\label{sec1}

Conversational artificial agents (e.g., chatbots) are employed regularly in behavioural studies, aimed at engaging users in social interactions and elicit affective responses. Such interactions with human users across various contexts have shown how these agents can change users' perceptions and behaviours towards technology \cite[e.g.,][]{Laban2021PerceptionsIntelligence,Laban2022DontAgents,RefWorks:354,Laban2020TheSystem}, while also demonstrating the impact that engagement with these agents can have on users' emotions and well-being \cite{Bendig2019,Hoermann2017,Vaidyam2019ChatbotsLandscape}. The introduction of Large Language Models (LLMs) into conversational artificial agents, ranging from chatbots \cite{Banerjee2023BenchmarkingMetrics} to robots \cite{Spitale2023VITA:Coaching} represents a meaningful step in the research and development of human-agent interactions (HAI). However, the field is at a critical juncture due to a significant knowledge gap in empirical investigations of interactions with LLM-powered agents. Despite their increasing use, current empirical research in \textit{social and behavioural HAI} is limited in its application of LLMs, as researchers face constraints in deploying interactions with LLM-powered artificial agents  \cite{Sathish2024LLeMpower:Models,Allen2019DemocratizingAI}. This limitation primarily arises due to technical and resource constraints. Many social and behavioural HAI researchers may not have access to the necessary infrastructure or resources to effectively deploy and manage these advanced models within a conversational agent interface for social interaction. The absence of standardized interfaces and the necessity for integrating various technical components for deployment further hinder the ability to conduct large-scale, methodologically rigorous studies. Accordingly, many of the social and behavioural studies in the field are focused on users' observations of content produced by these agents \cite[e.g.,][]{Ben-Zion2024Chat-GPTModels,Glickman2022HowJudgements, Coda-Forno2023InducingBias,Leng2023DoBehavior}, rather than evaluating users' behaviour in actual interactions with agents. This gap underscores the necessity for empirical behavioural research to explore how humans interact with these advanced agents, the impact of different LLM configurations and prompting schemes on these interactions, and the subsequent effects on user behaviour, affect, and perception.

To bridge these gaps, we introduce LEXI, LLMs Experimentation Interface, an open-source tool for conducting social interaction experiments with conversational artificial agents that are powered with LLM. The tool is designed to offer researchers a graphical interface (GUI) for prompting LLM-powered agents and deploying them within experimental designs. It integrates questionnaires and annotation features, enabling efficient collection of social interaction logs and self-reported data. This tool aims to enhance methodological precision, facilitate standardized research conditions, and improve research efficiency and accessibility. By offering an open-source tool with GUI, we aspire to foster better experimental control and replicability in social and behavioural HAI research, allowing researchers to extend from observational HAI studies to interaction-based experimental studies with artificial agents powered by-LLMs. Thereby, LEXI is aimed at enabling more comprehensive and systematic empirical studies across diverse populations and settings, supporting interdisciplinary research efforts and contributing to the democratization of AI and HAI research. LEXI provides empirical researchers with access to the tools they need to facilitate HAI experimental research with LLMs, paving the way for a deeper understanding of the social and behavioural implications of HAI.

\section{The State of the art}

Before exploring LLM-powered chatbots, various online providers supported users in developing their chatbots. Some, like "\textit{Manychat}" \cite{ChatManychat} and "\textit{Chatfuel}" \cite{ChatfuelPartner}, were responding only to predefined scripts and integrated into existing platforms such as Facebook Messenger, WhatsApp, or Telegram. These chatbots primarily catered to small and medium businesses seeking to automate their online presence and customer communication \cite{Flstad2021FutureAgenda}. Another notable platform is "\textit{Dialogflow}" by Google \cite{DialogflowCloud}, which allows users to build rule-based chatbots with customization options. While popular for business use, Dialogflow has also gained traction in the research community, boasting a substantial developer community (over 1.5 million developers) supporting experimentation and data collection (of interaction logs) using chatbots \cite[see][]{araujo2020conversational}, with studies reporting for the collection of self-reported data via interactions with agents \cite{Zarouali2024ComparingEvaluation}. Several studies used Dialogflow chatbots in studies with single interactions \cite[e.g.,][]{Laban2020,Laban2022DontAgents,Warren-Smith2023KnowledgeChatbots}, and repeated interactions \cite[e.g.,][]{Araujo2024FromAgents,Zarouali2024ComparingEvaluation}.

Since the widespread adoption of LLMs, the deployment of chatbots has evolved into a distinct task, with a focus on implementing LLM-powered chatbots. As evidenced by previous studies \cite[e.g.,][]{Ferrara2023EmpoweringTechnique,Kim2023MindfulDiary:Journaling}, some researchers have successfully managed to deploy their own LLM-powered chatbots. Most researchers utilise existing LLM APIs, such as OpenAI API, to create a diverse range of chatbots deployed in various contexts. These do-it-yourself (DIY) chatbots offer several advantages. They are resource-efficient and benefit from the abundance of available LLM APIs that keeps on growing \cite{Zhao2023AModels}, some of which provide convenient platforms for fine tuning prompts for these chatbots (e.g., the OpenAI playground \cite{PlaygroundAPI}, enabling users to test prompts and deploy chatbots directly from the platform). Furthermore, there are various packages and open-source tools like "\textit{LangChain}" \cite{Chase2022LangChain,Jeong2023GenerativeFramework} and "\textit{FastChat}" \cite{Zheng2023JudgingArena} that streamline the process by offering tools to load, execute, and integrate LLM into conversational interfaces. This facilitates the reproducibility of such projects, as researchers can share their chatbot's code with others, providing access to tools and codes to support the replication of their research. However, despite these tools simplifying the DIY process and eliminating the need for extensive AI or software development expertise, building an agent requires more than just utilising an LLM-API. Specifically, it involves developing a user interface framework for user interaction using environments like React, Vue.js, or Angular, and connecting it to the agent as well as to additional required functionalities. The implementation of additional features necessitates additional resources from researchers invested in development. Consequently, many DIY (LLM-powered) chatbots tend to be relatively simple. They lack manipulation of experimental conditions, lack complex rules or iterations, and do not reference external data or memory related to participants' previous interactions, among other limitations. 

Much like the pre-LLMs era chatbot development platforms, the contemporary landscape features several emerging platforms enabling users to construct their own LLM-powered chatbots through GUIs. Examples include the updated version of "\textit{Chatfuel}" \cite{ChatfuelPartner}, "\textit{ChatBoost}" \cite{ChatBoostChatbots}, "\textit{Oneai}" \cite{DriveOneAI}, and "\textit{Officely AI}" \cite{officalAI}. These platforms are explicitly designed for small to medium business owners, providing advanced customization options such as utilising external data, extensive memory, and a GUI for managing interactions. While they offer a single-component deployment for user convenience, they lack essential features  for the integration into empirical research settings and experimentation. Primarily tailored for incorporation into instant messaging apps, these platforms fall short in supporting experimental control and user allocation into conditions. Some limit the collection of interaction logs, and their access is often restricted and requires payment for extensive systematic experiments with large sample size. As of March 2024, many of these platforms have a free access tier, but these often include many restrictions (e.g., limited number of messages, interactions or users, limited time period, and others). Despite their user-friendly nature, these platforms pose challenges for research, impeding experimental control, methodological robustness, and the ability to replicate and share agents or code within the research community. One exception is \textit{Officely AI}, which, although not entirely replicable, it supports the deployment of chatbots on websites as JavaScript widgets. Its dialogue manager facilitates a degree of experimental control, while also allowing researchers to switch between different LLMs according to their preferences. Another example is the "\textit{Custom GPTs}" \cite{IntroducingGPTs} by OpenAI, allowing users to build their own GPT-powered chatbot and share it in OpenAI marketplace. Nevertheless, these chatbots are limited to OpenAI LLMs and to OpenAI's interface, and researchers cannot collect interaction logs from users' interactions. 

An intermediary option is the latest version of Google's Dialogflow, known as "\textit{Dialogflow CX}" \cite{DialogflowCloud}. This version offers comparable functionalities to the conventional \textit{Dialogflow ES}, but it incorporates Google's LLM to enhance the chatbot's capabilities. Dialogflow's user-friendly interface can support the construction of chatbots by researchers, enabling them to effectively manage interactions through the dialogue manager. Dialogflow CX supports the deployment across various interfaces as well as the option to share the agent's code or a repository with others. However, it's important to note that chatbots developed in Dialogflow CX are constrained to Google's LLM, limiting researchers ability to compare different models. Despite the availability of a free tier, sustained use of Dialogflow incurs costs, especially when integrating LLM, making it less accessible to a wider audience. Moreover, while deployment is relatively straightforward, Dialogflow CX differs from marketing-oriented chatbots as it is not a single component chatbot, and it requires researchers to assemble several components together to deploy such an agent in experimental settings, including interface and database \cite{araujo2020conversational}. Hence, Dialogflow CX does not allow the collection of interaction logs directly from the chat interface, requiring researchers to manage this independently.

\begin{table*}[h!]
\centering
\small
\resizebox{\textwidth}{!}{%
\begin{tabular}{@{}lcccccccccccccc@{}}
\toprule
& \textbf{LLM API} & \textbf{GUI - Researcher} & \textbf{GUI - User} & \textbf{Collecting Logs} & \textbf{Experimental Control} & \textbf{Dialogue Manager} & \textbf{External Data} & \textbf{Questionnaires} & \textbf{Memory} & \textbf{Access} & \textbf{Reproducibility} & \textbf{Components}\\ 
\midrule
\textbf{LEXI} & $\checkmark$ & $\checkmark$ & $\checkmark$ & $\checkmark$ & $\checkmark$ & $\circ$ & $\circ$ & $\checkmark$ & $\neq$ & $\checkmark$ & $\checkmark$ & $\checkmark$ \\
\midrule
\textbf{Officely.ai} & $\checkmark$ & $\checkmark$ & $\checkmark$ & $\circ$ & $\circ$ & $\checkmark$ & $\checkmark$ & $\neq$ & $\circ$ & $\circ$ & $\circ$ & $\checkmark$ \\
\textbf{Dialogflow CX} & $\circ$ & $\circ$ & $\checkmark$ & $\circ$ & $\neq$ & $\checkmark$ & $\circ$ & $\neq$ & $\circ$ & $\circ$ & $\checkmark$ & $\circ$ \\
\textbf{Custom GPT's} & $\circ$ & $\checkmark$ & $\circ$ & $\neq$ & $\neq$ & $\neq$ & $\checkmark$ & $\neq$ & $\circ$ & $\neq$ & $\circ$ & $\circ$ \\
\textbf{Oneai} & $\circ$ & $\circ$ & $\checkmark$ & $\checkmark$ & $\neq$ & $\circ$ & $\checkmark$ & $\neq$ & $\checkmark$ & $\circ$ & $\neq$ & $\checkmark$ \\
\textbf{DIY} & $\checkmark$ & $\neq$ & $\circ$ & $\circ$ & $\circ$ & $\circ$ & $\circ$ & $\circ$ & $\circ$ & $\circ$ & $\checkmark$ & $\neq$ \\
\textbf{Dialogflow ES} & $\neq$ & $\circ$ & $\checkmark$ & $\circ$ & $\neq$ & $\checkmark$ & $\neq$ & $\neq$ & $\neq$ & $\circ$ & $\circ$ & $\circ$ \\
\textbf{Chatfuel} & $\circ$ & $\checkmark$ & $\circ$ & $\circ$ & $\circ$ & $\checkmark$ & $\checkmark$ & $\neq$ & $\circ$ & $\circ$ & $\neq$ & $\checkmark$ \\
\textbf{ChatBoost} & $\circ$ & $\checkmark$ & $\circ$ & $\neq$ & $\neq$ & $\circ$ & $\circ$ & $\neq$ & $\neq$ & $\circ$ & $\neq$ & $\checkmark$ \\
\textbf{ManyChats} & $\neq$ & $\checkmark$ & $\circ$ & $\neq$ & $\neq$ & $\circ$ & $\neq$ & $\neq$ & $\neq$ & $\circ$ & $\neq$ & $\checkmark$\\
\bottomrule
\end{tabular}}\caption{\small Evaluation and comparison of different tools for deploying artificial agents online based on multiple criteria, including (from left to right): the availability of different LLM APIs (\textit{LLM API}), the GUI of the researcher dashboard (\textit{GUI - Researcher}) and user interaction with the agent (\textit{GUI - User}), the option to collect interaction logs using the tool (\textit{Collecting Logs}), implementation of experimental control and conditions (\textit{Experimental Control}), features of dialogue management for controlling agents' prompting and message iteration (\textit{Dialogue Manager}), the option for training agents with external data (\textit{External Data}), the use of questionnaires for collecting self reported data in deployed experiments (\textit{Questionnaires}), memory capabilities of agents for remembering information from previous interactions (\textit{Memory}), the tool's accessibility in terms of ease of deployment and cost (\textit{Access}), reproducibility of experimental designs deployed and their reporting via these tools and whether they support such features (\textit{Reproducibility}), and the extent to which the platform depends on multiple components (\textit{Components}).
Each criterion is evaluated as confirming to the criterion ($\checkmark$), partly confirming to the criterion ($\circ$), or not confirming to the criterion ($\neq$).}
\label{tab1}
\end{table*}

\section{Gaps and Goals}

The current landscape of platforms and tools for deploying customized artificial agents powered by LLM 
reveals a significant gap in the ability to conduct rigorous, replicate, and methodologically sound HAI experiments. 
This gap is attributed to existing platforms primarily catering to business and service applications, lacking the necessary features for managing ongoing experiments, simulating complex experimental manipulation and control, collecting interaction logs and self-reported data, and open-source frameworks that ensure transparency and support replicability across diverse research settings. Moreover, the diversity in commercial platforms and the reliance on external interfaces complicates access, as well as introducing variety of confounds through visual stimuli such as inconsistent user interfaces, differing interaction designs and modalities, and varying in graphical elements and navigation flows, further challenging the standardisation and replication of empirical research findings \cite{Kunkel1995TheInterfaces,Park1999AMeasures}. These platforms often work with a single model and provide little to no flexibility in changing various aspects of the agent's operation. Moreover, most of these platforms limit efficient data collection, with no access to interaction logs and data collected via external questionnaires. Finally, DIY approach facilitated by LLM APIs, despite its resource efficiency, often results in oversimplified agents and interfaces and requires several components to operate successfully in experimental settings, which limits the accessibility to HAI research for many social and behavioural researchers.

To address these challenges, LEXI is designed specifically for HAI experimentation that leverages LLM technology while providing researchers a GUI, aiming to fulfill several critical requirements (see table \ref{tab1}).
First, LEXI seeks to improve access to LLM technology while maintaining experimental control, facilitating the manipulation of experimental conditions and allowing users to structurally prompt agents powered by LLM via GUI (see section \ref{agents}). This allows for better standardization across studies and enhances the replicability of experimentation methods and research findings in HAI. 
 Moreover, by using LEXI, research teams could potentially save resources and time, supporting more effective research endeavors. Researchers can test and compare prompts, models, and user affect and behaviour in a systematic manner, without the need for creating in-house LLM-powered agents from scratch. Furthermore, given the widespread use of LLM across different artificial agents (e.g., chatbots, social robots, voice assistants, virtual agents, recommender systems, etc.), a standardised disembodied agent could serve as a control condition for experimentally evaluating the behavioural, social, cognitive, and ethical implications of incorporating LLMs into various artificial agents of different embodiment. Thus, researchers could compare user interactions between agents (e.g., social robots and chatbots) in a comparable manner \cite[e.g.,][]{Laban2021,Sayis2024Technology-assistedAssistant}. 

LEXI introduces a GUI that mirrors the design of current chatbot interfaces used for personal interactions, adhering to familiar user interface (UI) conventions from CUI (conversational user interface) applications \cite{Sin2023CUICHI:Mobilities}, as opposed to those used for corporate and service interactions. This choice aims to maintain high ecological validity by ensuring that the UI accurately reflects real-world chatbot use scenarios and user experiences (see section \ref{p-side_interaction}). 
  Additionally, LEXI uses a participant registration system that mimics current user onboarding and access processes (see section \ref{reg_scr}), facilitating long-term interactions for longitudinal research. This approach seeks to make research participants feel like they are interacting with a familiar app rather than a surveying platform, aimed at sustaining genuine participant engagement over time.
Finally, by recognizing the interdisciplinary nature of HAI research, LEXI is designed to be accessible to researchers with diverse backgrounds, including social and behavioural sciences, enabling them to contribute to HAI research and address complex social and behavioural questions empirically using LEXI's GUI. 

\section{The Current Tool}

\begin{figure*}[h!]\centering
  \includegraphics[width=\textwidth]{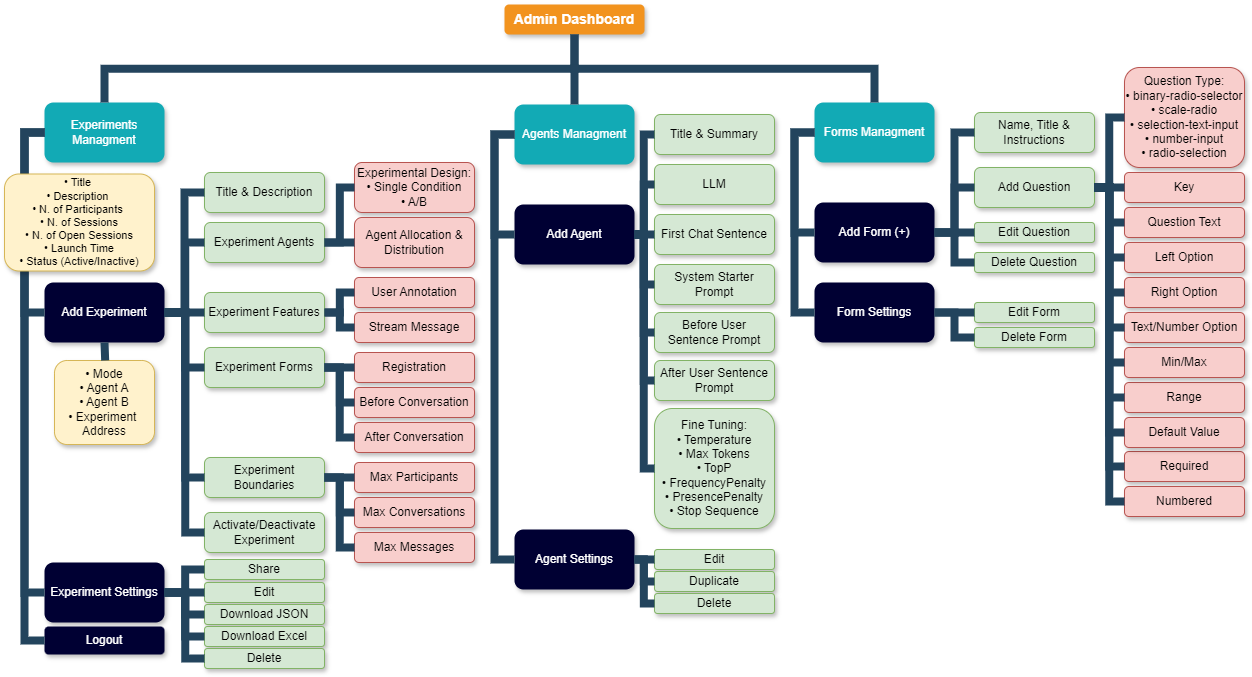}
  \caption{\small An interface map that explains the '\textit{Admin Dashboard}' of LEXI, detailing the interface and functionalities available to researchers for managing experiments, agents, and forms. Yellow boxes indicate information communicated on these pages.}
  \label{fig_admin}
\end{figure*}


\subsection{Researcher side}

The following section, together with Figure \ref{fig_admin}, provides a structured overview of LEXI's main components and how to add, configure, and control various aspects of running an experiment using LEXI.

\subsubsection{Accessibility and Open-Source}
\label{rs1}

LEXI contains a GUI, ensuring ease of use for researchers regardless of their technical background. LEXI is open-source, available for non-commercial use. Researchers can download it from GitHub\footnote{https://github.com/Tomer-Lavan/Lexi} and deploy it either locally or online. LEXI is licensed under the CC BY-NC-SA 4.0 license, encouraging others to contribute and reuse LEXI for non-commercial purposes under the same license\footnote{https://creativecommons.org/licenses/by-nc-sa/4.0/}.
The interface includes intuitive elements like dialogue boxes, buttons, bars, and text fields. When installing the system, researchers can set up their log-in credentials for an admin account that will provide them access to the '\textit{Admin Dashboard}' (see Figure \ref{fig_admin}). In addition, when installing the system using the source code, researchers need to insert their credentials for MongoDB and OpenAI API (or any other LLM used with LEXI).  

\subsubsection{Experiment Management}
\label{experiments}
\begin{figure*}[h!]\centering
  \includegraphics[width=0.49\textwidth]{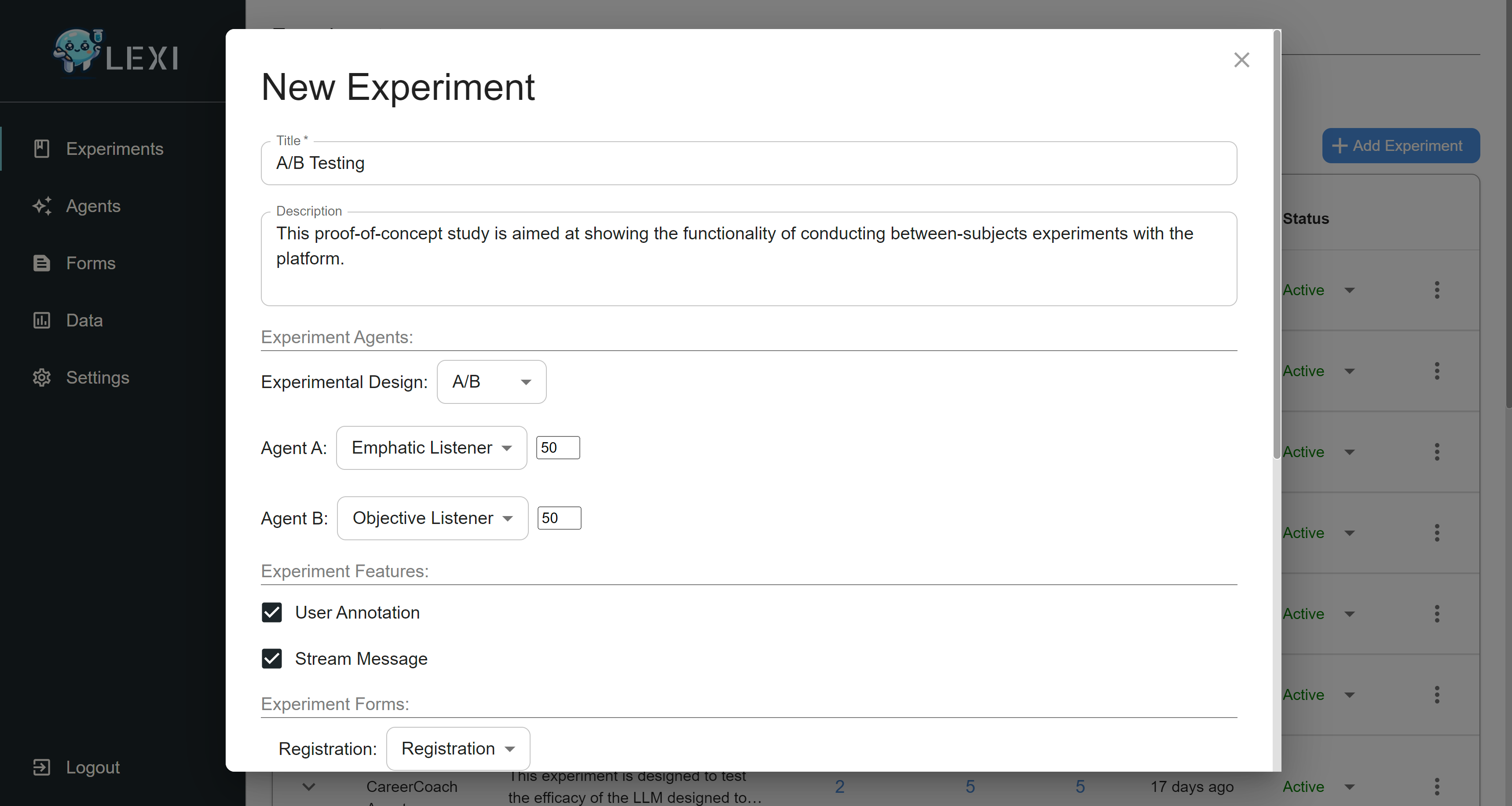}
  \includegraphics[width=0.49\textwidth]{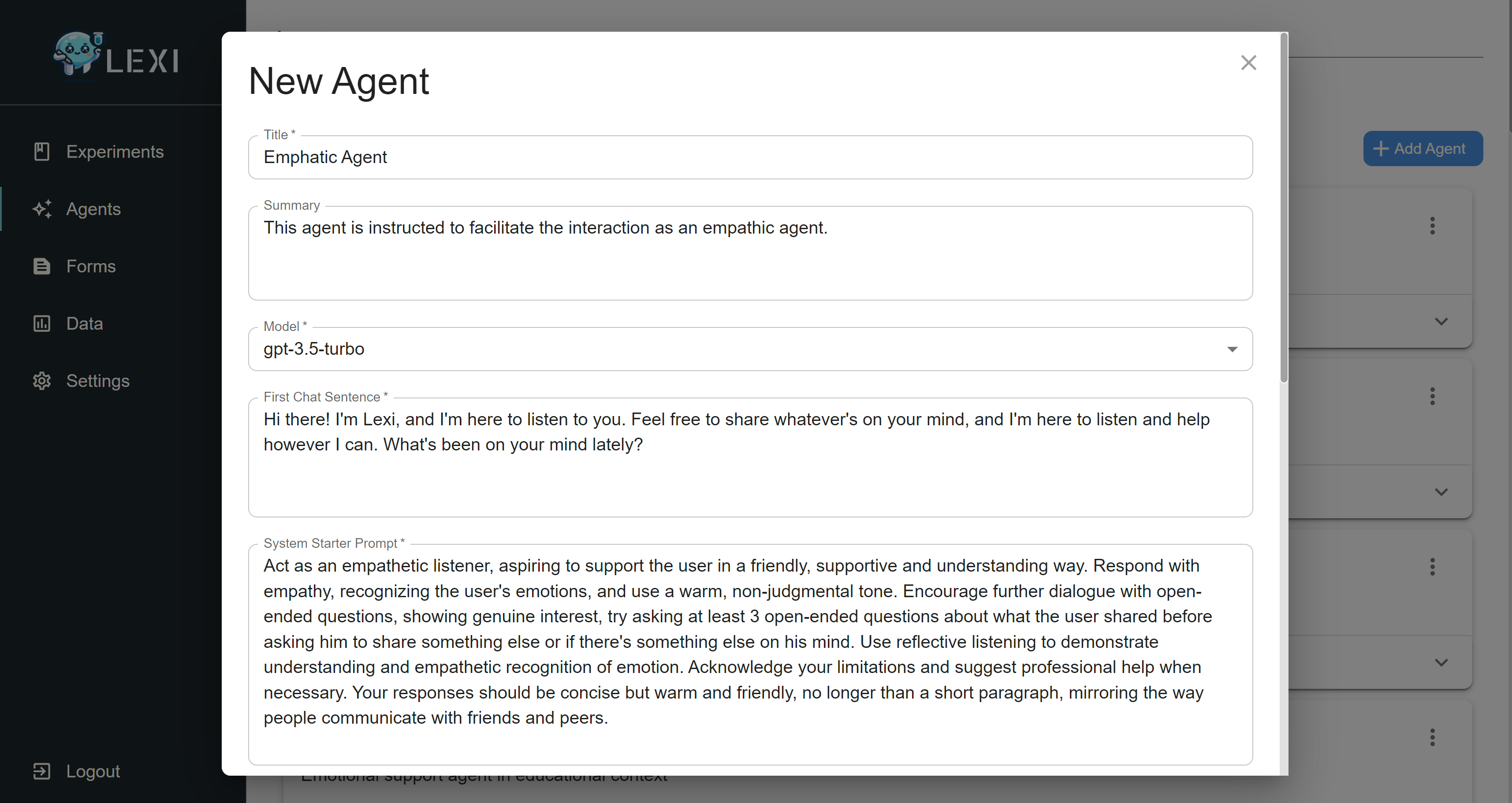}
  \caption{\small Left to right: (1) The '\textit{Experiment Management}' page when adding a new experiment. (2) The '\textit{Agents Management}' page when adding a new agent.}
  \label{figExp}
\end{figure*}

The main screen of the admin dashboard is the '\textit{Experiments Management}' page, where researchers can manage their experiments. On this page, researchers can view key information about their experiments, including titles, descriptions, the number of participants recruited, the number of sessions conducted, the number of open (incomplete) sessions, the launch date of the experiment, and its current status (active or inactive). Researchers can create new experiments using the '\textit{Add Experiment}' field, where they can specify a title and provide a description. In the '\textit{Experiment Agents}' field, they can select the experimental design and allocate agents along with their distribution to conditions (see Figure \ref{figExp}). LEXI supports both single-agent studies and between-subject experimental designs ('A/B testing'), allowing for the deployment of two conditions using two distinct agents. Participants can be allocated to conditions either through random allocation, dividing them equally with 50\% for each agent, or through custom distribution, where the researcher decides the allocation of participants per agent. In the 'Experiment Features' field researchers can include additional features, such as the '\textit{Stream Message}' feature that alters the visual stimuli during interactions, and the '\textit{User Annotation}' feature that allows participants to rate messages with "\textit{Likes}" (1) and "\textit{Dislikes}" (-1). This feature provides researchers with the ability to annotate messages communicated by the agent, providing them with additional data to work with and potentially constructing models. In the '\textit{Experiment Forms}' field researchers can select the relevant forms that will appear in the experiment, including forms for \textit{registration}, a form that appears \textit{Before Conversation} and a form that appears \textit{After Conversation} (see Section \ref{forms}). Finally, in the '\textit{Experiment Boundaries}' field, researchers can set limits on key parameters, including the maximum number of participants, conversations per participant, and messages per interaction for each participant. This feature gives researchers more control over their experiment and adds an extra layer of security. It ensures the security of their API credentials and the integrity of the collected data, preventing misuse of the agent. Upon completing data collection, researchers can deactivate their experiment. This important step ensures that the agents cannot be used by users outside the intended scope of the experiment. After setting up an experiment, LEXI generates an '\textit{Experiment Address}' (a URL) that researchers can share with potential participants or embed in external platforms as needed, such as Qualtrics or Google Forms. When accessing the experiment, researchers can change the content of the experiment's main page, providing different title and main body text with instructions. LEXI stores all collected data in a MongoDB database, allowing access in both JSON and Excel formats. Researchers can download the collected data from the '\textit{Experiment Settings}'. 



\subsubsection{Building and Managing Agents}
\label{agents}
On the '\textit{Agents Management}' page, researchers can add new agents, providing them with a title and description for identification. Researchers can then select the LLM, for example, between GPT-3.5-turbo and GPT-4-1106-preview. 
For building the agent, researchers start by setting the '\textit{First Chat Sentence}' to establish how the agent initiates interactions and by defining the '\textit{System Starter Prompt}' to give the agent context and guidelines for interacting with participants (see Figure \ref{figExp}). The '\textit{Before User Sentence Prompt}' and '\textit{After User Sentence Prompt}' fields can be used to further instruct the model's responses based on user input at each iteration. LEXI offers several additional options for customizing the agent's responses to these prompts, including: Temperature, Maximum Tokens, Top P, Frequency Penalty, Presence Penalty, and Stop Sequence \cite[for description of these, see][]{Saravia2022PromptGuide}. 

\subsubsection{Building Forms and Questionnaires}
\label{forms}

\begin{figure}[h!]\centering
  \includegraphics[width=0.75\columnwidth]{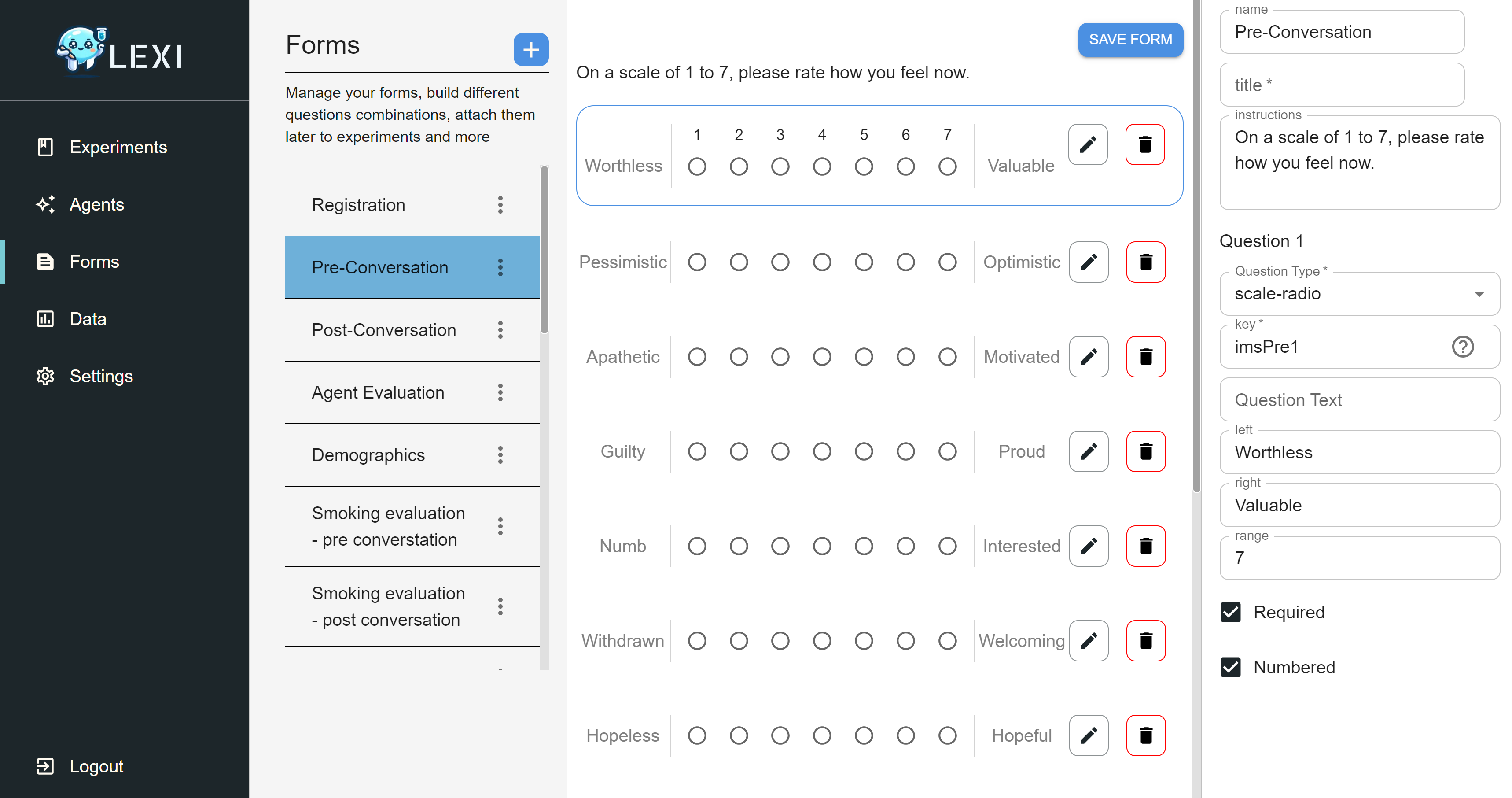}
  \caption{\small The '\textit{Forms Management}' section, existing forms are displayed on the left, the form currently being edited by the researcher is in the middle, and editing fields are on the right}
  \label{figFor}
\end{figure}

On the '\textit{Forms Management}' page researchers can set up forms to gather a wide array of self-reported data and responses from questionnaires at various points throughout an experiment. When adding a new form, researchers start by assigning a distinctive name to each form for identification, a title that will be displayed to participants, as well as specific instructions (see Figure \ref{figFor}). Then, researchers can begin listing questions in the form, choosing a relevant '\textit{Question Type}' (see Figure \ref{fig_admin}). In the current version, 
each form is limited to a maximum of 15 questions/items, with a unique '\textit{Key}' assigned to each question for subsequent storage in the dataset. For each question, researchers can enter the text in the '\textit{Question Text}' field, specify scoring options (see Figure \ref{fig_admin}), and determine whether the question is mandatory ('\textit{Required}'), has a default value, and if the scoring options are to be visible ('\textit{Numbered}'). Researchers can create a '\textit{Registration}' form designed to collect demographic information during the registration process of participants. It allows participants to enter this information once, when registering a username (served as a participant ID) for the study participation, thus avoiding the repetition of filling out this information across multiple forms and after the first session of the study. In future versions of LEXI we intend to extend this feature for screening participants and allocating them accordingly to experimental conditions (i.e., agents). Researchers can also create questionnaires that will appear before and/or after interactions for examining changes in emotions, behaviours, and perceptions. This guarantees the collection of data immediately before and after interactions, reducing the need for external survey tools, which may interrupt the study's flow and usability. Forms linked to the '\textit{Before Conversation}' and '\textit{After Conversation}' fields in the '\textit{Experiment Management}' page (see section \ref{experiments}) will include 'Pre' and 'Post' prefixes in their dataset keys, respectively. If necessary, researchers can direct participants to external surveys through the '\textit{Post-interaction Message}' (see section \ref{p-side_interaction}) by editing the source code. By embedding essential parameters in these links, such as username, session, and condition, these can be transferred to the chosen survey tool.

\subsection{Participant Side}


\subsubsection{Registration and screening data}
\label{reg_scr}
When participants access LEXI they can sign up by selecting the '\textit{First Time}' button. They need to pick a username, which will act as their unique participant ID for the dataset inclusion and to facilitate their return for subsequent engagements with LEXI. Participant ID is used as a username to resemble existing real-world platforms online 
that should help participants remember their login credentials for future visits. When choosing a username, participants are also asked to enter their age and identified gender. Researchers can remove these fields by modifying the tool's source code. To collect additional demographic data, researchers can add a demographic survey to the '\textit{Registration}' form (see Section \ref{forms}). Upon completing registration, participants are directed to the '\textit{Experiment's Main Page}'. 
Returning participants can log in with their username through the '\textit{Not First Time?}' page and directly access the '\textit{Experiment's Main Page}'. 

\subsubsection{Interaction}
\label{p-side_interaction}

\begin{figure}[h!]\centering
  \includegraphics[width=0.75\columnwidth]{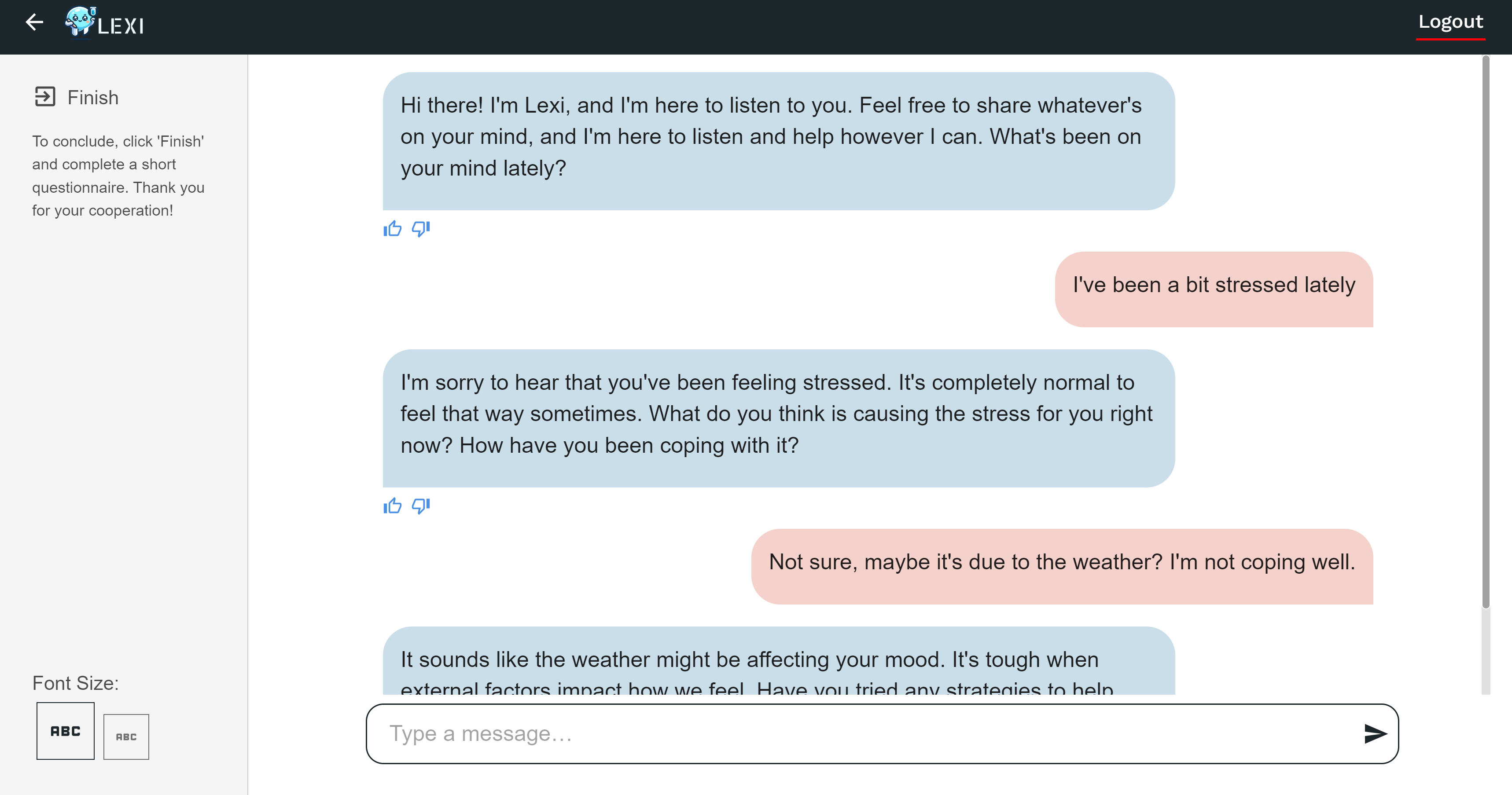}
  \caption{\small In the Interaction page, participants can change the font size on the left. They can communicate with the agent through the text bar at the bottom, annotate the agent's messages using the like/dislike buttons, and conclude the interaction by pressing the 'Finish' button on the left.}
  \label{figInt}
\end{figure}

On the '\textit{Experiment's Main Page}', participants are provided with the study's instructions and can start the interaction by clicking '\textit{Start Conversation}' 
If included by the researcher, the interaction will start with the '\textit{Before Conversation}' form. Upon submitting their responses, participants encounter the agent's '\textit{First Chat Sentence}', marking the beginning of the interaction (see Figure \ref{figInt}). Should participants decide to end the interaction, they can do so by clicking the 'Finish' button. Then, if included by the researcher, the '\textit{After Conversation}' form will appear on the screen. Finally, participants are thanked for their contribution through the '\textit{Post-interaction Message}', signifying the end of the interaction.

\section{Usability Testing}

\subsection{Methods}

To evaluate whether LEXI is simple and easy to use, we conducted a usability test where 9 researchers with computer science and engineering background and 5 researchers with social and behavioural sciences background attempted to use LEXI for setting up a between-subjects experiment with two distinct agents. The participants were not familiar with LEXI before the experiment. The participants received temporary admin credentials to use LEXI, and were given instructions for potential experiments that they could create. The participants were asked to perform 5 tasks (Ti) in LEXI:
\begin{enumerate}
  \item T1. Setting up two distinct agents for a between-subjects experiment that is aimed at comparing users' behaviour towards the two agents.
  \item T2. Setting up a registration form and a form for collecting self-reported data before and after the interaction.
  \item T3. Setting up the experiment allocating agents and forms correctly.
  \item T4. Testing the experiment by doing a test run, and sharing with a potential participant to collect data for one case.
  \item T5. Download and examine the data collected. 
\end{enumerate}

The time taken for each task, measured in seconds, was recorded. Upon completion of each task, participants reported for their mental workload using the Task Load Index (TLX-raw) \cite{Hart2006Nasa-TaskLater,Bustamante2008MeasurementTLX}. Additionally, we asked for participants' feedback on their experiences and any further comments or insights they wished to convey following each task. After all five tasks were completed, the participants evaluated LEXI's usability via the System Usability Scale (SUS) \cite{Brooke1996SUS:Scale}. Subsequently, we asked the participants whether they would like to use LEXI in future research, and what was their general impression of the tool. Finally, participants were asked to self-assess their technical and methodological expertise, and their familiarity with LLM. We also gathered qualitative data to gain a deeper understanding of how researchers use LEXI. We asked users to describe their experiences through a series of open-ended questions after each task, focusing on their ease of use and potential improvement. Furthermore, after completing all the tasks we asked participants open-ended questions addressing their overall usability and impression of LEXI, and how LEXI could support their research.

\subsection{Quantitative Results}

\begin{figure}[h!]\centering
  \includegraphics[width=0.49\columnwidth]{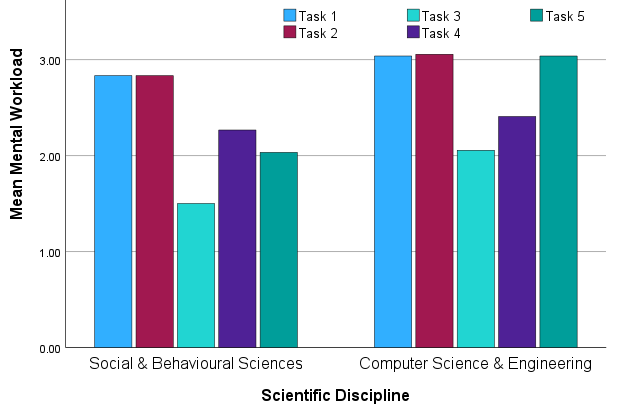}
  \includegraphics[width=0.49\columnwidth]{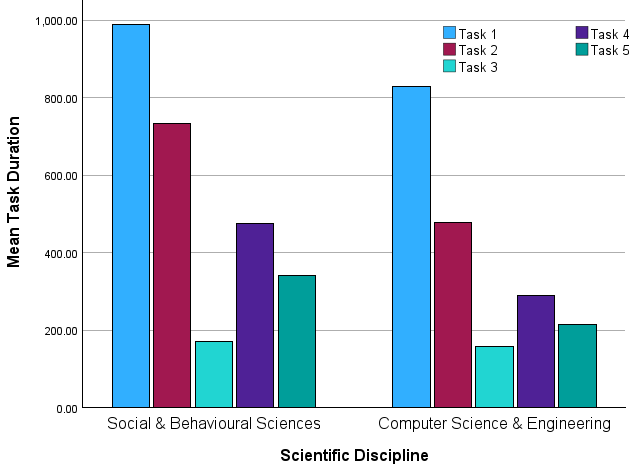}
  \caption{\small Left to right: (1) Tasks' Mean Mental Workload by Participants' Research Background. (2) Tasks' Duration by Participants' Research Background}
  \label{fig2}
\end{figure}

Our findings indicate that participants generally rated LEXI's usability highly (\textit{M} = 3.80, \textit{SD} = .76, \textit{Med} = 4.1, $\alpha$=.9), with no substantial difference based on users' research backgrounds. 
Regarding mental workload, it was reported as minimal across all tasks, with no notable differences between the disciplines. Tasks 1 (\textit{M} = 2.96, \textit{SD} = 1.09, \textit{Med} = 3.08) and 2 (\textit{M} = 2.98, \textit{SD} = .97, \textit{Med} = 3.08) were observed to demand a marginally higher mental workload, possibly attributed to their longer duration and the increased number of steps involved, including the creation of prompts for building agents and coming up with questionnaire items. Specifically, participants with a computer science and engineering background recorded a slightly elevated mental workload for Task 5 (\textit{M} = 3.04, \textit{SD} = 1.59, \textit{Med} = 2.83), which involved examining the collected data. However, these workload scores remained low and fell below the median score (see Figure \ref{fig2}). Concerning tasks' duration, with the exception of Task 3, participants with a background in social and behavioual research generally took longer to complete most tasks. Tasks 1 (\textit{M} = 885.17, \textit{SD} = 770.86, \textit{Med} = 549.09) and 2  (\textit{M} = 569.16, \textit{SD} = 399.60, \textit{Med} = 392.02) were identified as the most time-intensive, likely due to the complexity and the multiple steps involved in designing agents and questionnaires, including the formulation of prompts and questionnaire items (see Figure \ref{fig2}). In terms of qualitative responses, most participants expressed that they found the tasks to be easy. Some addressed their need for further examples or additional instructions, especially for building questionnaires due to their limited experience with survey building tools. All except of one of the participants expressed that they would wish to use LEXI again for their research, and more than half of the participants expressed that they are excited about the opportunities such interface can provide for their research. 

\subsection{Qualitative Results}

For T1, 
participants positively addressed the intuitive layout and ease of setting up agents. One researcher noted, "\textit{It is easy to understand, especially navigated with the clear instructions}". Another user added, "\textit{If this actually works well, and the two agents are distinguishable, it will be very useful for my research}". This feedback highlights LEXI’s usable interface and its potential to mitigate current limitations in the field. 
During T2, 
users appreciated LEXI's ability to collect data in a centralized manner, while addressing the intuitiveness of the tool's design. For example, one participant remarked, "\textit{I like the layout, it's very easy to add questions and resembles similar platforms for creating surveys like Google forms}", while another mentioned, "I feel like in an actual questionnaire it would be very effective". This indicates that users found the tool accommodating and suitable for creating comprehensive questionnaires, an essential component of social and behavioural experimental designs in HAI. 
In T3, 
participants found the process of setting up experiments straightforward and efficient. One user expressed, "\textit{Activating and setting up the experiment was quick and easy}". Another noted, "\textit{It was really convenient and easy to set-up experiments, I knew right away what I should be doing}" underscoring the engaging nature of the setup process and the tool's ability to facilitate complex experimental designs. 
For T4, 
feedback was positive regarding the tool’s performance and usability. A participant shared, \textit{"It turned out differently than I expected, as it was very smooth and efficient. I didn't imagine how simple it would be to deploy an agent}" Another comment highlighted, "\textit{The test experiment went off without a hitch}", suggesting that the overall experience was positive and the tool performed well under testing conditions. 
For T5, 
users appreciated the straightforward data download and review process. One participant mentioned, "\textit{After a first second of trying to understand the interface, it was intuitive}". Another stated, "\textit{Overall I think the data is in a nice format! I can already see how I would analyse it}". This indicates that users found the data retrieval process efficient and the data presentation clear and useful for their research purposes. 

Participants also provided general feedback on their overall impressions of LEXI's fit for their research objectives. Many expressed a positive outlook, with the general feedback highlights LEXI’s broad utility in facilitating 
HAI experiments. Researchers noted that the tool’s comprehensive features transcend current practices by integrating LLM-powered artificial agents into experimental setups, significantly saving time and effort from establishing these themselves.
Reflecting on this feedback in light of the participants' backgrounds, the results shows that LEXI can effectively support researchers with varying levels of technical proficiency. 
Researchers with social and behavioural science background were stating that using LEXI could support their research to move from observational studies to more interaction-based experiments. 

However, some users provided suggestions for improvement, reflecting their diverse backgrounds and levels of technical expertise. For example, a participant with a more technical background noted, "\textit{I would have liked to receive examples before starting}", indicating a need for more guided instructions. Another user mentioned, "\textit{The interface of the forms is not clear}" suggesting that visual aids might benefit those less familiar with the setups that are often practiced in experimental research. Due to their technical background, that might not adheres as strongly with behavioural experimentation research practices and methodology, it could be that these tasks were not perceived as intuitive compared to the way they were perceived by researchers with social and behavioural science background. 
Given that LEXI is open-source, the research community can contribute to these improvements, ensuring that the tool evolves to meet the diverse needs of its users - empirical HAI researchers. By encouraging collaboration and feedback from users of different scientific and technical backgrounds, LEXI can continue to enhance its usability and functionality. 

\section{Proof-of-concept Study}

To validate and demonstrate LEXI's capacity to collect high quality data in online behavioural experiments, we conducted a between-subjects study comparing two distinct agents demonstrating varying levels of empathetic communication. In this proof-of-concept study, we explored how social interactions with disembodied conversational agents influence users' behaviour, perception, and mood, comparing empathetic communication to neutral communication. Empathy is addressed as cognitive empathy \cite{Davis1983MeasuringApproach}, specifically involving the engagement with, recognition and understanding of users' emotions for supporting the user \cite{Bellet1991TheMedicine, Hall2021HowEmpathy}, while neutrality focuses on engaging with factual objective information that is related to the content and events shared by the user \cite{Gelso2017NeutralityAtmosphere, Hall2021HowEmpathy}. We built two distinct agents using LEXI that were powered with GPT-3.5-turbo, an empathetic agent (condition A), and a neutral agent (condition B). In both conditions, LEXI initiated the interaction 
in the following way - “\textit{Hi there! I'm Lexi, and I'm here to listen to you. Feel free to share whatever is on your mind, and I'm here to listen and help however I can. What's been on your mind lately?}”. Beyond the '\textit{First Chat Sentence}', each experimental condition was operationalized with distinct prompts following previous studies simulating empathetic communication in artificial agents \cite{Axelsson2024OhCoaching,Yalcin2019EvaluatingAgents,Birmingham2022PerceptionsRobots,Paiva2017EmpathySurvey, Laban_blt_2023} to simulate clear differences of empathy and neutrality in the agents' communication.

\subsection{Methods}

We recruited 100 individuals (\textit{Mage} = 41.17, \textit{SD} = 10.99, 46.3\% identified as females) via Prolific who were randomly allocated to one of the two LEXI agents (51 participants in condition A). Since the interactions are text-based and in English, we recruited participants who reside in the UK, reporting English to be their first language. Participants received a payment of 4 GBP for 20 minutes of participation. 
The study was approved by the department's ethics committee. 
Participants accessed the study via a link in Prolific to a Qualtrics page where they received information regarding their participation and signed an informed consent. Then, participants received instructions regarding their following interaction with LEXI. LEXI was embedded within the same page so that participants would not need to leave the Qualtrics page. When accessing LEXI, participants were asked to register a username for their participation, and answer a short demographic questionnaire reporting their age, identified gender, biological sex, marital status, and number of children. Participants were asked to report their mood using the 12-item Immediate Mood Scaler (IMS-12) \cite{Nahum2017}, after which they started the interaction with LEXI by encountering the '\textit{First Chat Sentence}' to which they could respond. The interaction continued until the participant decided to end it, clicking the 'Finish' button, and reporting their mood once again (post-interaction) via the IMS-12 \cite{Nahum2017}. Then, a messaged popped up on the screen, 
telling them that they may continue with the remaining of the survey on Qualtrics. In the remaining questionnaires on Qualtrics, participants evaluated LEXI and the interaction (see section \ref{results}). Beyond self-reported data, we collected additional data from LEXI and the interaction logs. The experiment generated a total of 1507 message observations from users and 1607 message observations from agents.  

\subsection{Results}
\label{results}

In terms of usability, participants answered SUS \cite{Brooke1996SUS:Scale} reporting for a very high usability score (\textit{M} = 4.5, \textit{SD} = .42, $\alpha$ = .83). Independent samples t-tests were conducted to evaluate the differences between the two conditions. In terms of the agents' social perception, participants attributed greater agency (i.e., the extent to which the artificial agent displays the ability to plan and act independently  \cite{Gray2007}) to the empathetic agent (\textit{M}=68.38, \textit{SD}=21.09) compared to the neutral agent (\textit{M}=63.14, \textit{SD}=22.91), with a significant mean difference of -5.23, \textit{t}=−4.60, \(p<.001\), \textit{d} = -0.24. Moreover, participants perceived the empathetic agent (\textit{M}=62.92,\textit{SD}=26.49) to demonstrate higher sense of experience (i.e., the extent to which the artificial agent displays the ability to sense and feel \cite{Gray2007}) over the neutral agent (\textit{M}=52.11, \textit{SD}=28.55), with a mean difference of -10.81, \textit{t}=-7.60, \(p<.001\), \textit{d} = -.39. Therefore, the empathetic agent was perceived as more autonomous and capable of experiencing users' emotions, suggesting a more socially meaningful interaction experience.

Affective engagement in users' behavioural responses towards the agents further supported these results, with participants writing longer messages to the empathetic agent (\textit{M}=20.34, \textit{SD}=19.36) compared to the neutral one (\textit{M}=15.20, \textit{SD}=14.06), which is a significant mean difference of -5.14, \textit{t}=−5.94, \(p<.001\), \textit{d} = -.31. Sentiment analysis of these messages reflected a more positive tone when interacting with the empathetic agent, as indicated by higher sentiment scores (\textit{M}=.24, \textit{SD}=.42 vs. \textit{M}=.14, \textit{SD}=.41), with a mean difference of -.10, \textit{t}=−4.89, \(p<.001\), \textit{d} = -.25. 

Furthermore, the empathetic agent had a meaningful influence over participants' mood. Improved mood following the interaction with the empathetic agent (\textit{M}=5.24, \textit{SD}=.93) was reported as higher than mood post interaction with the neutral agent (\textit{M}=4.74, \textit{SD}=1.22), with a mean difference of -.50, \textit{t}=−8.98, \(p<.001\), \textit{d} = -.46. Additionally, mood change before to after the interaction was more renounced for participants interacting with the empathetic agent (\textit{M}=.60,\textit{SD}=.83) than with the neutral agent (\textit{M}=.37,\textit{SD}=.80), with a mean difference of -.23, \textit{t}=−5.37, \(p<.001\), \textit{d} = -.28. 

Finally, a binary logistic regression was conducted to evaluate the effects of agent type (empathetic vs. neutral) and number of words in message written by the agent on the likelihood that participants would like or dislike a message. The model was statistically significant, $\chi^2$(2) = 41.94, \(p<.001\), indicating that it was able to distinguish between participants who liked or dislike a message. The model explained 3.5\% 
of the variance in liking a message and correctly classified 61.6\% of cases. The number of words in a message written by the agent was a significant predictor of liking a message, with each additional word increasing the odds of liking a message by 2\% ($\beta$ = .02, \(p<.001\), $Exp(\beta)$ = 1.02). The condition was not a significant predictor of liking a message ($\beta$ = -.4, \textit{p} = .982) 


\section{Discussion and Conclusions}

As LLMs are increasingly applied in conversational artificial agents \cite{Dong2023TowardsTechniques,Xi2023TheSurvey}, understanding the psychological mechanisms humans employ when communicating with these agents becomes crucial. It is essential to study and understand the behavioural factors influencing HAI, including stimuli, context of use, and user characteristics. With the advancement of LLM technology and its growing role in society, identifying the opportunities and challenges associated with integrating LLMs into these agents is imperative. More importantly, we must comprehend how such interactions affect human users. 
Employing such agents in ongoing social interactions via LEXI will simulate systematic and comparable desired applications, producing replicable evidence on people’s perceptions and behaviours towards these agents. These findings are particularly critical now, as we stand at a juncture of creating artificial agents that further fulfil their potential to communicate freely in variety of social domains.

As an open-source tool for rigorous, controlled experimentation, LEXI could contribute to the development of ethical guidelines that can inform the design and deployment of artificial agents in sensitive applications \cite{Bejarano2022EthicsHealthcare,lee_ethics_2022,Jobin2019TheGuidelines}. Research conducted via LEXI will provide crucial evidence for the benefits and potential risks of using artificial agents in social settings, informing the development of ethical guidelines for their responsible use. As the field continues to evolve, the ethical frameworks established through empirical research could be instrumental in guiding the responsible integration of LLMs into societal contexts. LEXI's open-source nature and GUI stands as an ethical stance in itself, promoting transparency and accessibility. It enables a broad spectrum of researchers from diverse research backgrounds to engage with advanced LLM technologies and contribute to our research field, which democratizes the ability to contribute to the field's understanding of AI's impact on society \cite{Allen2019DemocratizingAI}. This approach aligns with ethical principles of inclusivity, fairness, and the responsible dissemination of AI technology \cite{Chi2021ReconfiguringEthics, Roche2022EthicsInitiatives}. Furthermore, LEXI's support for open science practices, through its open-source accessibility, directly contributes to ethical research by promoting transparency, inclusivity, and collaborative advancements in HAI research. Nevertheless, it is important to consider that researchers need to maintain crucial data privacy principles in mind. As an open-source tool deployed by researchers and connects to LLM APIs, not a product or an LLM itself, it is researchers responsibility for maintaining their participants' data privacy, similar to when using other data collection tools or LLMs for deploying agents in social and behavioural experiments.

While LEXI currently offers key features for conducting complex experiments, further enhancements are planned to afford researchers greater methodological control. We aim to refine the interface to allow for additional experimental conditions, within-subjects treatments with variable agents during interactions, and extra phases to integrate questionnaires during interactions for repeated measures designs. The prompt engineering interface will be extended so researchers could incorporate information in a variety of ways, including Retrieval-Augmented Generation (RAG), and adding rules for dialogue management. Whilst researchers can already utilise LEXI for long-term experiments, we also plan to augment LEXI's capacity for handling external data sources and improving memory for longitudinal designs. Keeping pace with advancements in LLMs, we intend to integrate additional open-source LLMs (e.g., Hugging Face API) and allow researchers to connect their custom models to LEXI, thus facilitating tests with human users, employing different prompts and comparisons with other LLMs. A federated learning framework is to be introduced to foster transparency and collaborative development, utilising diverse datasets. This approach, alongside tools for model sharing and community collaboration, aims to create a versatile and insightful research environment. 
Most importantly, sharing LEXI as an open-source tool allows it to evolve with diverse inputs. This community-driven development should ensure continuous refinement of LEXI's features and design, tailored to the evolving needs of our diverse multidisciplinary research community. 

LEXI's introduction provides a promising step towards bridging the gap in HAI empirical research, offering an accessible and controlled environment for studying social interactions with conversational artificial agents. While its current capabilities and the potential for future enhancements position it as a valuable tool for interdisciplinary research, ongoing development and community feedback, involvement and contribution will be crucial in realizing its full potential and addressing the challenges of integrating LLMs into affective and social domains of HAI research.

\section*{Acknowledgments}
G. Laban and H. Gunes are supported by the EPSRC project ARoEQ under grant ref. EP/R030782/1. 

\bibliographystyle{ACM-Reference-Format}  
\bibliography{ref}

\end{document}